%% file: main.tex
\documentclass[conference]{IEEEtran}
\IEEEoverridecommandlockouts
\usepackage{cite}
\usepackage{amsmath,amssymb,amsfonts}
\usepackage{algorithmic}
\usepackage{graphicx}
\usepackage{textcomp}
\usepackage{xcolor}
\usepackage{listings}
\usepackage{xurl}
\usepackage[hidelinks]{hyperref}
\usepackage{todonotes}[]
\usepackage{booktabs}
\usepackage{tikz}
\usepackage{tcolorbox}
\usepackage{fbox}
\usepackage{multirow}
\usepackage[flushleft]{threeparttable}

\definecolor{lightgray}{rgb}{.9,.9,.9}
\definecolor{darkgray}{rgb}{.4,.4,.4}
\definecolor{purple}{rgb}{0.65, 0.12, 0.82}
\definecolor{light-gray}{gray}{0.80}
\lstdefinelanguage{JavaScript}{
  keywords={typeof, new, true, false, catch, function, return, null, catch, switch, var, if, in, while, do, else, case, break},
  keywordstyle=\color{blue}\bfseries,
  ndkeywords={class, export, boolean, throw, implements, import, this},
  ndkeywordstyle=\color{darkgray}\bfseries,
  identifierstyle=\color{black},
  sensitive=false,
  comment=[l]{//},
  morecomment=[s]{/*}{*/},
  commentstyle=\color{purple}\ttfamily,
  stringstyle=\color{red}\ttfamily,
  morestring=[b]',
  morestring=[b]"
}

\makeatletter
\def\footnoterule{\kern-3\p@
  \hrule \@width 2in \kern 2.6\p@} 
\makeatother

\def\BibTeX{{\rm B\kern-.05em{\sc i\kern-.025em b}\kern-.08em
    T\kern-.1667em\lower.7ex\hbox{E}\kern-.125emX}}

\tikzstyle{process} = [rectangle,  rounded corners, minimum width=2cm, minimum height=1cm, text centered, draw=black, fill=gray!30]
\tikzstyle{arrow} = [thick,->,>=stealth]

\newtcolorbox{qoutebox}[3][]
{
  colframe=black!30!white,
  colback  = #2!10,
  #1,
}

\begin{document}

 \title{An Empirical Study of Flaky Tests in JavaScript}

\author{\IEEEauthorblockN{Negar Hashemi\IEEEauthorrefmark{1},
Amjed Tahir\IEEEauthorrefmark{2} and Shawn Rasheed\IEEEauthorrefmark{3}}
\IEEEauthorblockA{Massey University\\
Palmerston North, New Zealand\\
\IEEEauthorrefmark{1}negar.hashemi.1@uni.massey.ac.nz,
\IEEEauthorrefmark{2}a.tahir@massey.ac.nz,
\IEEEauthorrefmark{3}s.rasheed@massey.ac.nz}}
\maketitle

\maketitle

\begin{abstract}
Flaky tests (tests with non-deterministic outcomes) can be problematic for testing efficiency and software reliability. Flaky tests in test suites can also significantly delay software releases.
There have been several studies that attempt to quantify the impact of test flakiness in different programming languages (e.g., Java and Python) and application domains (e.g., mobile and GUI-based).
In this paper, we conduct an empirical study of the state of flaky tests in JavaScript. We investigate two aspects of flaky tests in JavaScript projects: the main causes of flaky tests in these projects and common fixing strategies. By analysing 452 commits from large, top-scoring JavaScript projects from GitHub, we found that flakiness caused by concurrency-related issues (e.g., async wait, race conditions or deadlocks) is the most dominant reason for test flakiness. The other top causes of flaky tests are operating system-specific (e.g., features that work on specific OS or OS versions) and network stability (e.g., internet  availability or bad socket management). In terms of how flaky tests are dealt with, the majority of those flaky tests ($>$80\%) are fixed to eliminate flaky behaviour and developers sometimes  skip, quarantine or remove flaky tests.

\end{abstract}


\input{introduction}
\input{related-work}
\input{methodology}
\input{results}

\input{threats}
\input{conclusion}

\section*{Acknowledgment}
This work was partially funded by a SfTI National Science Challenge grant No. MAUX2004.

\bibliographystyle{IEEEtran}
\bibliography{main}

\end{document}

%% file: introduction.tex
\section{Introduction}
\label{sec:intro}

Test flakiness is a significant issue that affects the quality of software products. The impact of test flakiness is most apparent in regression testing, as regression tests rely on tests having deterministic outcomes to ensure that code changes do not break existing functionality. Flaky tests, which have non-deterministic outcomes due to factors such as concurrency, randomness, shared state and reliance on external resources, can make tests unreliable for this purpose. Such flaky behaviour is problematic as it leads to intermittent breaks in builds, uncertainty in choosing corrective measures for failing tests and bug fixes. This can have a negative impact on development, product quality, and  delivery \cite{fowler2011eradicating,SandhuTesting2015,palmer2019}. Flakiness also impacts testing-related activities such as test suite minimization \cite{machalica2019predictive} and test parallelization \cite{lam2020dependent}, as well as techniques that rely on testing such as fault localization \cite{vancsics2020} or program repair \cite{qin2021}.

Flaky tests are not only problematic, but they are also quite common in large codebases. For example, it was reported that around 16\% of tests at Google are flaky, and 1 in 7 of the tests occasionally fail in a way that is not caused by changes to the code or tests \cite{googleFlaky2016}, making it a real challenge for automated testing \cite{googleFlakyAutomated2020}. Vahabzadeh et al. \cite{vahabzadeh2015empirical} found that 21\% of the false alarms in Apache projects are caused by flaky tests. It was also shown that around 13\% of failed builds in CI pipelines are due to flaky tests \cite{labuschagne2017measuring}.

With the increased attention given to flaky tests in research and practice, there have been a number of studies that investigated different aspects of flaky tests, such as detection or elimination techniques. Previous studies on test flakiness and its causes largely focus on specific sources of test flakiness, such as order-dependency \cite{gambi2018practical}, concurrency \cite{dong2020concurrencyrelated} or UI-specific flakiness \cite{memon2013automated,romano2021empirical}. There have been some empirical studies that investigated flakiness in the context of specific programming languages. These studies aim to understand the prevalence of test flakiness, underlying causes and strategies employed by developers to respond to them. The first empirical study by Luo et al. \cite{luo2014empirical} focusses on flakiness fixing commits mined from Apache projects (mostly written in Java). A similar approach was followed for Android applications in Thorve et al. \cite{Thorve2018empirical}. A study on flakiness in Python applications \cite{sjobom2019studying} examines existing tests (by running the tests themselves) rather than mining commits that fix flaky tests.

To the best of our knowledge, there are no studies that specifically investigate flaky test causes (beyond UI-causes \cite{romano2021empirical}) in JavaScript projects, despite its popularity and extensive use. JavaScript has been the most commonly used programming language for several years
\footnote{\url{https://insights.stackoverflow.com/survey/2021\#technology-most-popular-technologies}}.
It is also popular in various application domains, including web (server-side and client-side), mobile, and the Internet of Things (IoT) \cite{javascript2018popular,lima2021exposing}. In addition, given the nature of the language (handling HTTP requests, running on browsers), JavaScript is known to be  one of the languages that uses asynchronous APIs extensively, with an execution model susceptible to races. Asynchronization and other concurrency bugs are known sources of flakiness in other languages, in particular Java and Python.

In this paper, we present the first extensive empirical study of flaky tests in JavaScript.
In particular, the goal of this paper is to investigate how prevalent flaky tests are in JavaScript projects, and to understand the causes and impact of flaky tests in these projects. To further investigate flaky tests, we collect and analyse 452 commits from popular JavaScript projects on GitHub. For each flaky test, we inspect the commit messages, the pull requests, and the changes to the code to understand the cause of flakiness. Also, we identify the main strategies that developers follow to deal with flaky tests.

The remainder of this paper is structured as follows: we present related work on flaky tests in Section \ref{relatedlab}. Our research questions and methodology are explained in Section \ref{methodologylab}. We present our detailed results and answer our research questions, followed by a discussion of the results in Section \ref{resultlab}. Finally, we present our conclusion in Section \ref{conclusionlab}.

%% file: related-work.tex
\section{Related Work}
\label{relatedlab}


Several studies published in recent years investigated the main causes and impact of flaky tests in both open-source and proprietary software.
In a study of the maintenance of the regression test suites in GitHub projects \cite{labuschagne2017measuring}, it was observed that about 13\% of the test failures are due to flaky tests.
A survey with developers on the precipitation of flaky tests by Ahmad et al. \cite{ahmad2019empirical} identified 23 factors that are perceived to affect test flakiness (including technical and organisational factors). 
Similarly, Eck et al. \cite{eck2019understanding} studied developers’ precipitation when it comes to flaky tests (including Mozilla projects' developers) to examine the nature and the origin of 200 flaky tests that had been fixed by the same developers. 
It was found that flakiness is perceived as a significant issue by the vast majority of developers surveyed. 

Lam et al. \cite{Lam2020Longitudinal} conducted a large-scale longitudinal study of flaky tests to determine when those tests become flaky, and what changes cause them to become so. 
They found that 75\% of the flaky tests (184 out of 245) are flaky when added, and 85\% of flaky tests can be detected when detectors are run on newly added or directly modified tests. 
Vahabzadeh et al. \cite{vahabzadeh2015empirical} studied the types of bugs in test code and observed that flaky tests and semantic bugs constitute the dominant cause of tests provoking false alarms in the test code.

There have been a few empirical studies on flaky tests in different programming languages and application domains, but none we could find that studied flakiness, as an issue in itself, specifically in JavaScript. We discuss a number of those studies below, and provide a summary of flaky test causes in Table \ref{tabcauseref}.

Luo et al.\cite{luo2014empirical} reported the first extensive empirical study of flaky tests, categorizing causes and fixing strategies of flaky tests by studying 201 commits of fixes in open-source Apache projects. The root causes of flakiness were split into ten main categories: async wait, concurrency, test order dependency, resource leak, network, time, IO, randomness, floating point operations, and unordered collections.



Gruber et al. \cite{gruber2021empirical} analysed projects from the Python Package Index\footnote{\url{https://pypi.org/}} to study the cause of the detected flaky tests, and found that order dependency is the main cause of flakiness in Python projects. The study identified \textit{infrastructure flakiness} as a new type of test flakiness that has not been reported previously.
Lam et al. \cite{lam2020study} studied the lifecycle of flaky tests in six large-scale projects at Microsoft, and found that asynchronous calls are the leading cause of flaky tests in those projects. They also proposed an automated solution, called Flakiness and Time Balancer (FaTB), to reduce the frequency of flaky-test failures caused by asynchronous calls.

The impact of flaky tests has also been studied in different application domains. A study of flaky tests in Android apps \cite{Thorve2018empirical} identified two new causes to those discussed in \cite{luo2014empirical}, namely: program logic and UI. Dutta et al. \cite{Dutta2020Detecting} studied flaky tests in applications that use probabilistic programming or machine learning frameworks, finding that \textit{randomness}, as expected, is the biggest cause of flakiness in those applications. 
Romano et al. \cite{romano2021empirical} analysed 235 flaky UI tests from 62 web and Android projects, and identified four categories of causes for UI-based flaky tests.
Moran et al. \cite{Mor2020FlakyLoc} focused on detecting the root cause of flakiness in web applications by analysing test execution under different combinations of the environmental factors that may trigger flakiness. 

There have also been a number of 
tools that have been targeted to detect certain types of flaky tests such as \textit{DeFlaker} \cite{bell2018deflaker}, \textit{RootFinder} \cite{Lam2019Root}, \textit{iFixFlakies} \cite{Shi2019iFixFlakies}, \textit{SHAKER} \cite{Silva2020Shake} and \textit{FlakeScanner} \cite{Dong2021Concurrency}.

\begin{table*}[htp]
\caption{Causes of flaky tests as identified in prior studies}\label{tab: }
\label{tabcauseref}
\begin{center}
\begin{tabular}{llll}
\toprule
\textbf{Luo et al.}\cite{luo2014empirical} & \textbf{Thorve et al.}\cite{Thorve2018empirical} & \textbf{Dutta et al.} \cite{Dutta2020Detecting} & \textbf{Gruber et al.}\cite{gruber2021empirical}
 \\
\hline\\[-0.5em]
Async Wait & Concurrency & Algorithmic Non-determinism & Infrastructure\\
Concurrency & Dependency & Floating-point Computations & Test Order Dependency\\
Test Order Dependency & Program Logic & Incorrect/Flaky API Usage & Network\\
Resource Leak & Network & Unsynced Seeds & Randomness\\
Network & UI & Concurrency & IO\\
Time & ~ & Hardware &Time\\
IO & ~ & Other & Async Wait\\
Randomness & ~ & ~ &Concurrency\\
Floating Point Operations & ~ & ~ & Resource leak\\
Unordered Collections & ~ & ~ &Test Case Timeout\\
~ & ~ & ~ &Unordered Collections\\

\bottomrule
\end{tabular}
\end{center}
\end{table*}

%% file: methodology.tex
\section{Study Methodology}
\label{methodologylab}

The main goal of this study is to investigate causes and strategies followed to deal with flaky tests in JavaScript. In this paper we answer the following two research questions: \\

\noindent \textbf{RQ1} What are the main causes of flaky tests in JavaScript projects?\\


\noindent \textbf{RQ2} What are the strategies followed when dealing with flaky tests in JavaScript projects?\\

Our target is to analyse commits that are believed to refer to tests with flaky behaviour, rather than exposing flakiness by compiling and analysing the programs themselves. We conducted the study in two phases. We first collect commits (including commit messages and source code changes) from open-source JavaScript projects and then manually analyse the causes of flakiness as noted in these commits. We also analyse the strategies followed when those flaky tests were fixed.

\subsection{Dataset}\label{ds}
 We constructed a dataset of JavaScript projects obtained from GitHub. We targeted only popular projects based on the number of stars of each project (based on the \textit{star} rating on GitHub). GitHub's star feature allows users to mark their interest or helpful repositories on GitHub. In a way, it is a metric that shows how much interest a project has drawn.
We found that the number of stars is highly correlated with the number of forks and contributors, indicating popularity.

To search through the repositories, we used a GitHub API\footnote{\url{https://docs.github.com/en/rest/reference/search}} to search through all publicly available GitHub repositories and filter results of the commits for analysis. We first extract the top 40 starred JavaScript projects on GitHub. We choose the top 40 projects because they provide a large enough sample (number of commits) that is deemed suitable for our analysis. 
We compared the number of commits we retrieved by the ones in similar previous studies:  Luo et al. \cite{luo2014empirical} retrieved 486 commits in total, but analysed 201 commits that fix flaky tests; while
Thorve et al. \cite{Thorve2018empirical} retrieved 77 commits. Romano et al. \cite{romano2021empirical} analysed 235 flaky UI test. In comparison, we extracted a total of 316,948 commits and ended up with 735 flaky tests' related commits.

\begin{table*}[t]
\caption{Projects from the top 40 most starred JavaScript projects on GitHub with commit messages containing at least one of the search keywords}
\label{tab:repo}
\begin{center}

\begin{tabular}{lllp{0.06\linewidth}p{0.1\linewidth}llp{0.23\linewidth}}

\hline\\[-0.5em]
\textbf{Repositories} &\textbf{\#Stars} &\textbf{\#Commits}&\textbf{JavaScript File} &\textbf{\#Flakiness Related Commits} &\textbf{\#Contributor} &\textbf{Age} &\textbf{Description} \\
\hline\\[-0.5em]
freeCodeCamp/freeCodeCamp & 320k & 28253 & 47.4\%& 4 & 4287 & Oct. 2014 & Corpus of educational JavaScript projects \\
vuejs/vue & 188k & 3200 & 97.6\%& 2 & 388 & Dec. 2013 & JavaScript framework for building UI \\
facebook/react & 174k & 14448 & 95.7\%& 22 & 1495 & July 2013 & JavaScript library for building user interfaces\\
twbs/bootstrap & 153k & 21165 & 41.4\%& 5 & 1231 & Oct. 2011 & Web development JavaScript framework \\
electron/electron & 96.6k & 25574 & 6.5\%& 48 & 1069 & July 2013 & Desktop apps framework \\
nodejs/node & 81.6k & 34487 & 61.2\%& 467 & 3012 & May 2009 & Cross-platform JavaScript runtime environment\\
denoland/deno & 77.7k & 6310 & 23.6\%& 3 & 635 & May 2018 & A JavaScript and TypeScript runtime environment\\
mrdoob/three.js & 74.6k & 38172 & 54.2\%& 1 & 1482 & Apr. 2010 & JavaScript 3D Library\\
mui-org/material-ui & 70.9k & 17626 & 58.2\%& 19 & 2278 & Nov. 2014 & A framework to build React apps\\
storybookjs/storybook & 64.5k & 35707 & 9\%& 14 & 1365 & April 2016 & UI components development tool\\
atom/atom & 56k & 38404 & 88.3\%& 68 & 488 & Dec. 2013 &  Cross-platform text editor\\
jquery/jquery & 55.3k & 6536 & 93.6\%& 9 & 276 & Jan. 2006 & JavaScript Library\\
chartjs/Chart.js & 54.7k & 4043 & 98.3\%& 1 & 380 & June 2014 & Data visualization library \\
ElemeFE/element & 50.8k & 4500 & 26.7\%& 1 & 556 & Sep. 2016 & A Vue.js-based UI Toolkit for Web\\
lodash/lodash & 50.6k & 8005 & 100\%& 2 & 310 & April 2012 & A modern JavaScript utility library  \\
moment/moment & 46k & 3961 & 99.7\%& 8 & 590 & Jan. 2015 & A JavaScript date library \\
meteor/meteor & 42.6k & 24211 & 92.3\%& 50 & 465 & Jan. 2012 & JavaScript web framework \\
yarnpkg/yarn & 40.1k & 2346 & 98.7\%& 11 & 521 & June 2016 & JavaScript package manager \\
\hline\\[-0.5em]
\textbf{Total}&&316948& &735 &20828 &&\\
\hline

\end{tabular}
\end{center}
\end{table*}


\subsection{Mining Process}
\begin {figure*}[htp]
\centering
\includegraphics[scale=0.8]{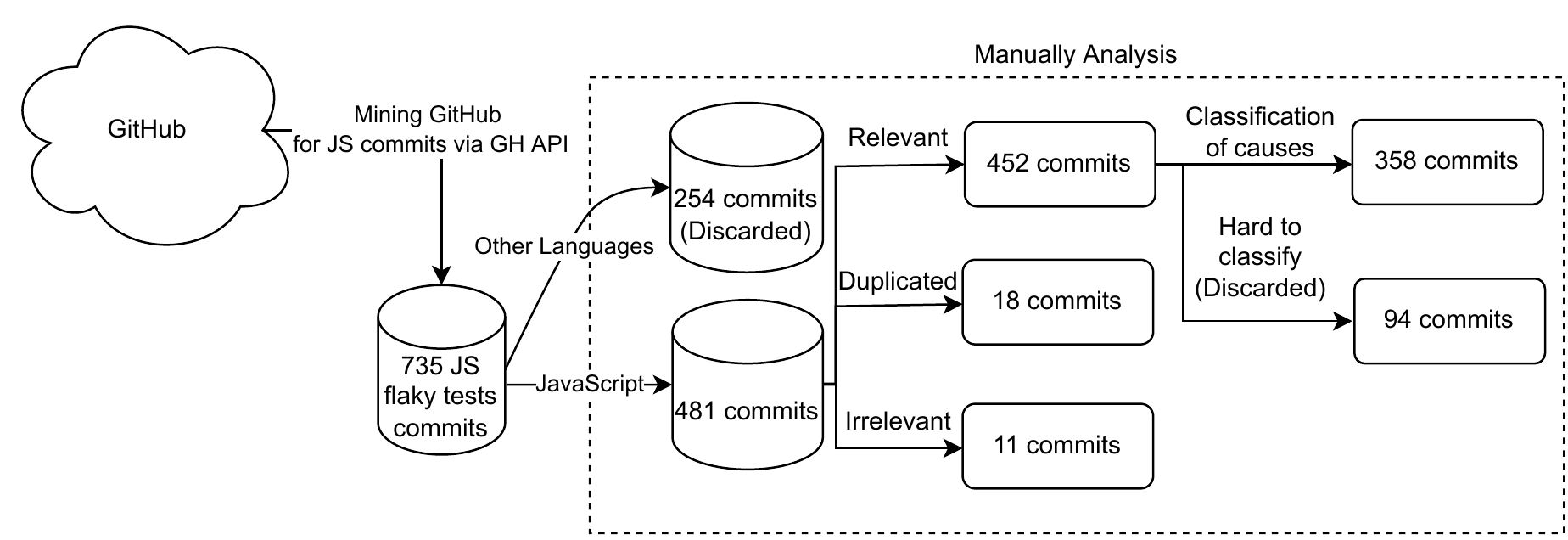}
\caption{Mining flaky tests' commits from GitHub's JavaScript projects}
\label{fig:mining}
\end{figure*}

 Our mining process is shown in Fig. \ref{fig:mining}. By using GitHub API, we searched through 40 repositories for the following keywords:  ``flaky'', ``flakey'', ``flakiness'', ``intermit'', ``fragile'', ``brittle'', ``Intermittent'', and ``non-deterministic test''.  
We found 18 repositories to have at least one commit with one of the above keywords we used. This resulted in a total of 735 commits. 
 We conducted this search on May 8th, 2021. Table \ref{tab:repo} shows statistics from the projects we analysed.

 Out of those 735 commits, we found that 254 of them are related to languages other than JavaScript, so we excluded those as we are interested only in flakiness that is related to JavaScript. This left us with a total of 481 commits that are relevant to our analysis.
%
%
%
%
%
%

\subsection{Manual Analysis of Commits}
We manually analysed all 481 commits that we included. For each commit, we analysed the commit message, any code changes in the files in the repository as well as the associated issues (if any) in the issue tracker. 
We started our process by manually filtering out commits related to languages other than JavaScript (out of 735 commits, we found 481 to be JavaScript-related). Two of the authors then individually and separately classified all 481 commits based on the cause of flakiness and then compared their classification results. To inform our classification, we checked the associated issues (if any), pull requests and changes in the test and CUT. When there were classification disagreements between the two authors, a third co-author was then involved to resolve conflicts. The third author analysed a total of 139 commits (where the two authors did not agree or were not sure). Disagreements between all authors were settled through discussions until a 100\% agreement was reached.

We categorize each of these commits into one of the following :
\begin{itemize}
  \item \textbf{Relevant}: commits that are directly related to flaky tests.
  \item \textbf{Irrelevant}: commits that contain one of the keywords, but it is not related to flakiness.
  \item \textbf{Duplicated}: identical commits.
\end{itemize}


Of those 481 commits, we found that 11 commits were found to be irrelevant (not related to test flakiness), 
and 18 commits were duplicated, thus those were excluded. This resulted in a total of 452 commits that we considered relevant, which we manually classified.
We provide a full replication package (including scripts and data) online at: \url{https://doi.org/10.5281/zenodo.6757825}.

%% file: results.tex
\section{Results}
\label{resultlab}


\input{results-RQ1}
\input{results-RQ2}
\input{implications}

%% file: results-RQ1.tex
\subsection{RQ1: What are the main causes of flaky tests in JavaScript projects?}
\label{sec:results:RQ1}

To answer RQ1, we manually analysed and categorised all commits based on the relevant causes of flakiness (based on the commit message or the change in the files associated with the commit).
We used the list of causes noted in  previous empirical studies on test flakiness as our baseline (see Table \ref{tabcauseref}). We then classified each commit based on the list of flakiness causes. If the cause of flakiness is new or not in the list, we then add it as a new cause and continue with our classification.

During this process, we were not able to categorise or determine the cause of flakiness in 94 commits. Hence, those were categorised as \textit{``unsure''} or \textit{``hard to categorise''} and we continued to analyse the remaining 358 commits. 

As it turns out, most of the causes we identified fit well into one or more of the categories noted in previous studies (i.e., \cite{luo2014empirical,Thorve2018empirical,Dutta2020Detecting,gruber2021empirical}). However, some of the causes we identified resulted in new categories that best reflect the nature of flakiness and thus those have been categorised separately.

Table \ref{tab:result} shows the results of the top 10 causes we found in the JavaScript projects we analysed. We present the results of our categories below, together with examples of each cause from the projects we analysed.

\begin{table}[!htp]
\caption{Overall results of flaky test categories (causes) in JavaScript}
\label{tab:result}
\begin{center}
\begin{threeparttable}
\begin{tabular}{lrr}
\toprule

\textbf{Cause of Flakiness}  &\textbf{\# of Commits} & \%\\

\midrule
Concurrency  &74 &20.7\%\\
Async Wait  &70 &19.6\% \\
OS &66 &18.4\% \\
Network &45 &12.6\% \\
Platform  &37 &10.3\%\\
UI &21 &5.9\% \\
Hardware  &17 &4.7\%\\
Time  &12 &3.4\%\\
Resource Leak  &10 &2.8\%\\
Other &13 & 3.6\% \\
\midrule
\textbf{Total} & 365\tnote{*} & \\
\bottomrule
\end{tabular}
\begin{tablenotes}
  \item[*] The total number of commits is 358 as there are 7 commits related to 2 causes of flakiness.
\end{tablenotes}
\end{threeparttable}
\end{center}
\end{table}

\noindent \textbf{Concurrency} We classify a commit in this category when a test is flaky due to any concurrency-related issues in the test or code under test (CUT) such as
event races, atomicity violations or deadlocks \cite{wang2017comprehensive}. 
We found that 74 of the flaky tests belong to this category. For example, in this snippet\footnote{\url{https://tinyurl.com/2s32r7xk}} from Node.js, using \texttt{setImmediate} in line 6 could lead to race conditions, which lead to non-deterministic test outcomes.

\begin{lstlisting}
onst keepOpen = setTimeout(() => {
@@ -20,7 +20,7 @@ const timer = setInterval(() => {
    timer._onTimeout = () => {
      throw new Error('Unrefd interval fired after being cleared.');
    };
    setImmediate(() => clearTimeout(keepOpen));
  }
}, 1);
 \end{lstlisting}



Another example of a concurrency related flaky tests is the following example from Meteor\footnote{\url{https://tinyurl.com/ysde5t37}}. This test is flaky because the template rendered callbacks get called after flush time, but not if the template got destroyed. If the client managed to respond to the server rejecting the method before the client's flush cycle, the rendered callback would never fire.

\begin{lstlisting}[escapechar=@]
testAsyncMulti('spacebars - template - defer in rendered callbacks', [function (test, expect) {
  var tmpl = Template.spacebars_template_test_defer_in_rendered;
  var coll = new Meteor.Collection("test-defer-in-rendered--client-only");
  tmpl.items = function () {
    return coll.find();
  };
  var subtmpl = Template.spacebars_template_test_defer_ in_rendered_subtemplate;
  subtmpl.rendered = expect(function () {
    Meteor.defer(function () {
    });
  });
  var div = renderToDiv(tmpl);
  Meteor._suppress_log(1);
  coll.insert({});
}]);
\end{lstlisting}


\noindent \textbf{Async Wait} A commit is labelled to be in this category when a test makes an asynchronous call and does not wait properly for the result of the call to become available before using it. Async Wait is in fact a subcategory of concurrency, but we classify it separately here mainly because it represents nearly half (48\%) of the concurrency-related flaky tests. In the projects we analysed, there are 70 commits in total that fits into the async wait category, such as the following example from Electron\footnote{\url{https://tinyurl.com/2p8zr5k4}}:

\begin{lstlisting}
 it('should clear the navigation history', async () => {
    loadWebView(webview, {
    nodeintegration: 'on',
    src: `file://${fixtures}/pages/history.html`
  })
const event = await waitForEvent(webview, 'ipc-message')
 \end{lstlisting}


In this test, the IPC message was sent before it was fully loaded, so the history did not contain everything it should have, and cannot clear all the pages, leading to non-deterministic outcomes.

In the following test\footnote{\url{https://tinyurl.com/2p8ebjbx}} from FreeCodeCamp, using \texttt{cy.contains('Go to next challenge').click();} in line 10 and then not waiting for the result before visiting the next page can cause flakiness.

\begin{lstlisting}
describe('project submission', () => {
   it('Should be possible to submit Python projects', () => {
    const { superBlock, block, challenges } = projects;
    challenges.forEach(challenge => {
      cy.visit(`/learn/${superBlock}/${block}/${challenge}`);
      cy.get('#dynamic-front-end-form')
        .get('#solution')
        .type('https://repl.it/@camperbot/python-project#main.py');
      cy.contains("I've completed this challenge").click();
      cy.contains('Go to next challenge').click();
\end{lstlisting}


\noindent \textbf{Operating Systems (OS)} A commit is classified to be in the OS category when the test fails due to a run in a specific OS or OS version, while passing on others. It can happen because of a test dependency on specific OS features or environmental settings. We classify OS as a separate category of root causes from the platform (discussed later) because it represents a large percentage of flaky tests by itself. In total, we found that 66 commits belong to the OS category. The following test\footnote{\url{https://tinyurl.com/533ycmsc}} from Node.js is a good example of a flaky test in this category.

\begin{lstlisting}
if (process.platform === 'darwin') {
  setTimeout(function() {
    fs.writeFileSync(filepathOne, 'world');
  }, 100);
 } else {
  fs.writeFileSync(filepathOne, 'world');
}
 \end{lstlisting}


In OS X, events for \texttt{fs.watch()} might only start showing up after a delay. 
 To work around that, there is a timer in the test that delays the writing of the file by 100ms, but in some cases this might not be enough. So the event never fired, and the test times out.

Another example of an OS flaky test from Angular\footnote{\url{https://tinyurl.com/2p9d7rxk}} shows a test that resulted in a \texttt{ECONNREFUSED} error (i.e., refused connection) for only Windows, but it works well for other OS.
\begin{lstlisting}
function launchChromeAndRunLighthouse(url, flags, config) {
  const launcher = new ChromeLauncher({autoSelectChrome: true});
  return launcher.run().
    then(() => lighthouse(url, flags, config)).
    then(results => launcher.kill().then(() => results)).
    catch(err => launcher.kill().then(() => { throw err; }, () => { throw err; }));
}
\end{lstlisting}


\noindent \textbf{Network} A commit is  in this category when a test fails due to remote connection failures (e.g., lost internet connection when accessing an external URL) or local bad socket management. There are 45 commits that fit into the network category. One example we see from this category is the following test from Node.js\footnote{\url{https://tinyurl.com/yc6c97m6}}, which is using port `0' for an IPv6-only operation and assuming that the OS would supply a port that was also available in IPv4. However, the CI results seem to indicate that a port can be supplied that is in use by IPv4 but available to IPv6, resulting in the test failing.

\begin{lstlisting}
  net.createServer().listen({
    host,
    port: 0,
    ipv6Only: true,
  }, common.mustCall());
\end{lstlisting}

Another example from React\footnote{\url{https://tinyurl.com/yfprp6tx}} shows a test that uses absolute URLs, which can end up in flakiness. 

\begin{lstlisting}
var __dirname = __filename.split('/').reverse().slice(1).reverse().join('/');
window.ReactWebWorker_URL = __dirname + '/../src/test/worker.js' + cacheBust;
document.write('<script src="' + __dirname + '/../build/jasmine.js' + cacheBust + '"><\/script>');
document.write('<script src="' + __dirname + '/../build/react.js' + cacheBust + '"><\/script>');
document.write('<script src="' + __dirname + '/../build/react-test.js' + cacheBust + '"><\/script>');
document.write('<script src="' + __dirname + '/../node_modules/jasmine-tapreporter/src/tapreporter.js' + cacheBust + '"><\/script>');
document.write('<script src="' + __dirname + '/../test/the-files-to-test.generated.js' + cacheBust + '"><\/script>');
document.write('<script src="' + __dirname + '/../test/jasmine-execute.js' + cacheBust + '"><\/script>');
\end{lstlisting}


\noindent \textbf{Platform} The commits in this category include all flaky tests related to platform, e.g., test environment, browsers. We excluded OS related flaky tests here as those, although are platform related, are a large enough subcategory that we discussed separately earlier in this section.  
We identified 37 commits in total from the platform category. The following example\footnote{\url{https://tinyurl.com/3xms6muw}} from Electon shows a test that, due to some specifications, failed on one CI, but passed on another. The test times out regularly on Travis CI but pass when built on Jenkins.

\begin{lstlisting}
it('can be manually resized with setSize even when attribute is present', done => {
      w = new BrowserWindow({show: false, width: 200, height: 200})
      w.loadURL('file://' + fixtures + '/pages/webview-no-guest-resize.html')
 \end{lstlisting}


Similarly, the following test\footnote{\url{https://tinyurl.com/348bzef5}} from Node.js is flaky due to reliance on platform timing in lines 3, 5, 8, 9, 11, and 18.

\begin{lstlisting}
const common = require('../common');
const fs = require('fs');
const platformTimeout = common.platformTimeout;
const t1 = setInterval(() => {
  common.busyLoop(platformTimeout(12));
}, platformTimeout(10));
const t2 = setInterval(() => {
  common.busyLoop(platformTimeout(15));
}, platformTimeout(10));
const t3 =
  setTimeout(common.mustNotCall('eventloop blocked!'), platformTimeout(200));
setTimeout(function() {
  fs.stat('/dev/nonexistent', () => {
    clearInterval(t1);
    clearInterval(t2);
    clearTimeout(t3);
  });
}, platformTimeout(50));
\end{lstlisting}


\noindent \textbf{UI} The commits in this category include all flaky tests related to UI features, e.g., flakiness due to different windows display, the blinking cursor, or the highlight decoration. There are 21 tests that belong to the UI category. The following example\footnote{\url{https://tinyurl.com/yckkhh4a}} from Atom shows a test aims to verify that the cursor blinks when the editor is focused and the cursors are not moving. It expects the blinking cursor to start in the visible state, and then transition to the invisible state. But the initial state of the cursor is not important, since the cursor toggles between the visible and the invisible state. 

\begin{lstlisting}
await component.getNextUpdatePromise()
      const [cursor1, cursor2] = element.querySelectorAll('.cursor')
      expect(getComputedStyle(cursor1).opacity).toBe('1')
      expect(getComputedStyle(cursor2).opacity).toBe('1')
      await conditionPromise(() =>
        getComputedStyle(cursor1).opacity === '0' && getComputedStyle(cursor2).opacity === '0'
      )
      await conditionPromise(() =>
        getComputedStyle(cursor1).opacity === '1' && getComputedStyle(cursor2).opacity === '1'
      )
      await conditionPromise(() =>
        getComputedStyle(cursor1).opacity === '0' && getComputedStyle(cursor2).opacity === '0'
      )
 \end{lstlisting}


Also, the following Atom test\footnote{\url{https://tinyurl.com/mvdvm9vf}} is flaky due to UI related issues. Resize events are unreliable and may not be emitted right away. This could cause the test code to wait for an update promise that was unrelated to the resize event (e.g., cursor blinking).

\begin{lstlisting}
 setEditorHeightInLines(component, 13);
 await setEditorWidthInCharacters(component, 50);
 expect(component.getRenderedStartRow()).toBe(0);
 expect(component.getRenderedEndRow()).toBe(13);
\end{lstlisting}


\noindent \textbf{Hardware} This category combines all flakiness in test outcomes that are resulted from hardware-dependencies. We found a total of 17 commits related to the hardware category. For example, in the following example\footnote{\url{https://tinyurl.com/mttkmtx5}} from Node.js, the test runs in Raspberry Pi. The number of clients is set to 100, which has caused some flaky behaviour when run in Raspberry Pi. 

\begin{lstlisting}
var responses = 0;
var N = 10;
var M = 10;
server.listen(common.PORT, function() {
 for (var i = 0; i < N; i++) {
   setTimeout(function() {
     for (var j = 0; j < M; j++) {
       http.get({ port: common.PORT, path: '/' }, function(res) {
         console.log('%d %d', responses, res.statusCode);
         if (++responses == N * M) {
           console.error('Received all responses, closing server');
           server.close();
\end{lstlisting}


Another example from this category is the following test from Node.js test\footnote{\url{https://tinyurl.com/2rfs42zs}}, which, when looping rapidly and making new connections (in line 5) it can cause the
Raspberry Pi 2 Model to malfunction.

\begin{lstlisting}
const server = net.createServer(function listener(c) {
  connections.push(c);
}).listen(common.PORT, function makeConnections() {
  for (var i = 0; i < NUM; i++) {
    net.connect(common.PORT, function connected() {
      clientConnected(this);
    });
\end{lstlisting}

\noindent \textbf{Time} The commits in this category have flakiness that comes from time related issues, e.g., relying on the system's time. We classify 12  commits in total into the time category. The following example\footnote{\url{https://tinyurl.com/354h3uf4}} shows a flaky test that is the result of relying on the system's time with the use of \texttt{Date.now()} function.

\begin{lstlisting}
const now = Date.now();
while (now + 10 >= Date.now());
 \end{lstlisting}


The following snippet\footnote{\url{https://tinyurl.com/mvfmf8rz}} from Moment can cause flakiness because of checking the equality of times that can be different by a millisecond.

\begin{lstlisting}
test.expect(7);
test.equal(moment.utc().valueOf(), moment().valueOf(), "Calling moment.utc() should default to the current time");
\end{lstlisting}


\noindent \textbf{Resource Leak} We classify a commit in this category in cases where the CUT 
did not properly manage resources, e.g., memory allocations issues, database connections, loss of connection, or not enough space. There are 10 commits from the resource leak category. 
The following example\footnote{\url{https://tinyurl.com/2s44nnua}} from Node.js shows a flaky test due to \texttt{EMFILE}, which means that a process is trying to open too many files.

\begin{lstlisting}
if (accumulated.includes('Error:') && !finished) {
    assert(
      accumulated.includes('ENOSPC: System limit for number ' +
                           'of file watchers reached'),
      accumulated);
 \end{lstlisting}



The following snippet\footnote{\url{https://tinyurl.com/4yycbc73}} from Node.js shows a test that is not performing proper clean-up and so it would fail if run more than one time on the same machine.

\begin{lstlisting}
function test_up_multiple(cb) {
  console.error('test_up_multiple');
  if (skipSymlinks) {
    console.log('skipping symlink test (no privs)');
    return runNextTest();
  }
  fs.mkdirSync(tmp('a'), 0755);
  fs.mkdirSync(tmp('a/b'), 0755);
  fs.symlinkSync('..', tmp('a/d'), 'dir');
@@ -432,6 +443,7 @@ function test_up_multiple(cb) {
      if (er) throw er;
      assert.equal(abedabed_real, real);
      cb();
     });
  });
}
\end{lstlisting}

\noindent \textbf{Other Known Causes}
Other remaining causes (13 commits) are related to several categories. The remaining causes are (each provided with an example from the projects we analysed):  IO\footnote{\url{https://tinyurl.com/2zv3acya}}, test order dependency\footnote{\url{https://tinyurl.com/39nnnt85}}, floating point operations\footnote{\url{https://tinyurl.com/2p999bjw}}, randomness\footnote{\url{https://tinyurl.com/3utzckbb}}
, and implementation dependency\footnote{\url{https://tinyurl.com/yc2vtvzz}}.

In addition, we categorized 7 commits in total into more than one cause. For those commits, we believe that there are multiple causes that may have an equal effect on the flaky behaviour. For example, the following test\footnote{\url{https://tinyurl.com/2p8hwj7w}} from Node.js falsely assumed that closing the client (which also currently destroys the socket rather than shutting down the connection) would still leave enough time for the server side to receive the stream error.

\begin{lstlisting}
server.on('stream', common.mustCall((stream) => {
  stream.on('error', common.mustCall(() => {
  stream.on('close', common.mustCall(() => {
      server.close();
    }));
  req = client.request();
  req.resume();
  req.on('error', common.mustCall(() => {
    req.on('close', common.mustCall(() => {
      client.close();
\end{lstlisting}


Both the network and the wait for the connection issues have the same impact on the test being flaky. Hence, we categorized this test under \textit{network} and \textit{async wait} categories.

\begin{qoutebox}{white}{}
\textbf{RQ1 findings:} The top four causes of test flakiness in JavaScript projects are concurrency (21\%), async wait (20\%), OS (18\%) and network (13\%). Other causes of flaky tests we found include platform, UI, hardware, time and resource leak.
\end{qoutebox}

%% file: results-RQ2.tex
\subsection{RQ2: What are the common strategies followed when dealing with flaky tests in JavaScript projects?}
\label{sec:results:RQ2}

When developers face a flaky test, they usually fix it by changing the test code, the CUT, or both. We found that a total of 350 commits to fix flakiness by changing the test code, 3 commits changed CUT, and 5 commits made changes to both. Although those changes aim to fix the flaky tests, they do not necessarily completely get rid of the flaky behaviour, but instead they improve/reduce such behaviour. 
In other words, the changes sometimes decrease the chance of the flaky behaviour, but do not eliminate it.





\begin{figure}
\centering
\includegraphics[scale=0.8]{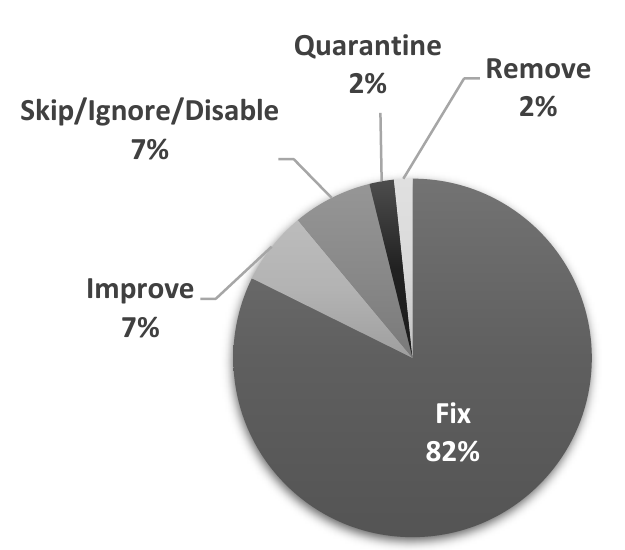}
\caption{Distribution of Response Strategies}
\label{fig:responce}
\end{figure}

For each commit, we identify the main strategies that developers use to deal with flaky tests. We categorise those into one of five different strategies, as follows. A visual summary of the distribution of commits based on the followed strategy is shown in Fig.~\ref{fig:responce}.

\noindent \textbf{Fix:} This strategy aims to fix any flaky test that has been detected/reported (after reproducing and confirming the flaky behaviour). We found that 295 commits to present an immediate fix of the noted flaky test. This is not surprising as we mainly mined commits with known flaky tests (that is, flaky tests that have already been detected by the developers, and thus it is very likely to have been actioned). The following example\footnote{\url{https://tinyurl.com/mttkmtx5}} from Node.js shows a flaky test from the \emph{Hardware} category (the test failed due to set numbers of clients in Raspberry Pi). 
The CI results\footnote{\url{https://tinyurl.com/2zz6fmvs}} show that it is fixed by reducing the number of clients from 100 to 16. 
\begin{lstlisting}[escapechar=@]
@\colorbox{light-gray}{var N = 10;}@
@\colorbox{light-gray}{var M = 10;}@
var  N = 4;
var M = 4;
server.listen(common.PORT, function() {
  for (var i = 0; i < N; i++) {
    setTimeout(function() {
      for (var j = 0; j < M; j++) {
        http.get({ port: common.PORT, path: '/' }, function(res) {
\end{lstlisting}


\noindent \textbf{Improve:} In this strategy, the test code or the CUT are changed to decrease the chance of flakiness or to make the code more traceable to fix it later (which makes it technical debt in nature). There are 23 commits that in this category. The following example\footnote{\url{https://tinyurl.com/nwmexy27}} from Atom shows an added delay to the tests to make them less flaky, but they did not fix the root problem (an explanation is provided in the comment section in the following listing).

\begin{lstlisting}
// In Windows64, in some situations nsfw (the currently default watcher)
// does not trigger the change events if they happen just after start watching,
// so we need to wait some time. More info: https://github.com/atom/atom/issues/19442
await timeoutPromise(300);
\end{lstlisting}


%
%

\noindent \textbf{Skip/ignore/disable:} This strategy provides an option to developers to \textit{skip} or \textit{ignore} the test that is flaky. In some cases, especially when the developer is fully aware of the implications of those flaky tests, they may decide to \textit{disable} those flaky tests and continue with the build as planned.
We found 26 commits in total that belong to this category. For example, in the following code\footnote{\url{https://tinyurl.com/bdzewfc7}} from Electron, there is a condition to skip the test for ``win32" platform or ``arm64" architecture.
\begin{lstlisting}
 if describe(process.platform !== 'win32' || process.arch !== 'arm64')('did-change-theme-color event', () => {
\end{lstlisting}

The developer noted the following in a pull request\footnote{\url{https://tinyurl.com/3epjpe5b}}:
\begin{quote}
\textit{``...there are a couple of tests that are consistently flaky on arm (both Linux and Windows variants). This PR disables those flakes so that we can get our ARM CI to the point where it can be relied on for PR validation instead of maintainers ignoring it."}
\end{quote}

\noindent \textbf{Quarantine:} In this strategy, developers isolate flaky tests from other healthy tests by keeping them in a quarantined area (e.g., different test suite or development branch) in order to diagnose and then fix those tests. This is a common strategy used in practice (e.g., \cite{googleFlaky2016,flexport2021,uber2021}). 
 There are 8 commits that belong to this category. For example, the code\footnote{\url{https://tinyurl.com/4ybhbhj7}} from Node.js move a portion of the test to a separate file 
as the test is flaky on CentOS. The developer then noted that:
\begin{quote}
\textit{``[test] has been flaky on CentOS. This allows us to .. eliminate the cause of the flakiness without affecting other unrelated portions of the test."}
\end{quote}


\noindent \textbf{Remove:} The strategy suggests that all flaky tests should be removed from the test suite. We found 6 commits that remove the test to completely eliminate the flaky behaviour. For example, there is a commit\footnote{\url{https://tinyurl.com/3uzke4rf}} from Node.js shows a tests that was completely removed the test from the test suit due to being flaky.
The developer noted the following:

\begin{quote}
\textit{``[the test] is supposed to test an internal debug feature, but what it effectively ends up testing, is the exact lifecycle of different kinds of internal handles ... making the test fail intermittently ... It's not a good test, delete it."}
\end{quote}

\begin {figure*}
\centering
\includegraphics[width=\textwidth]{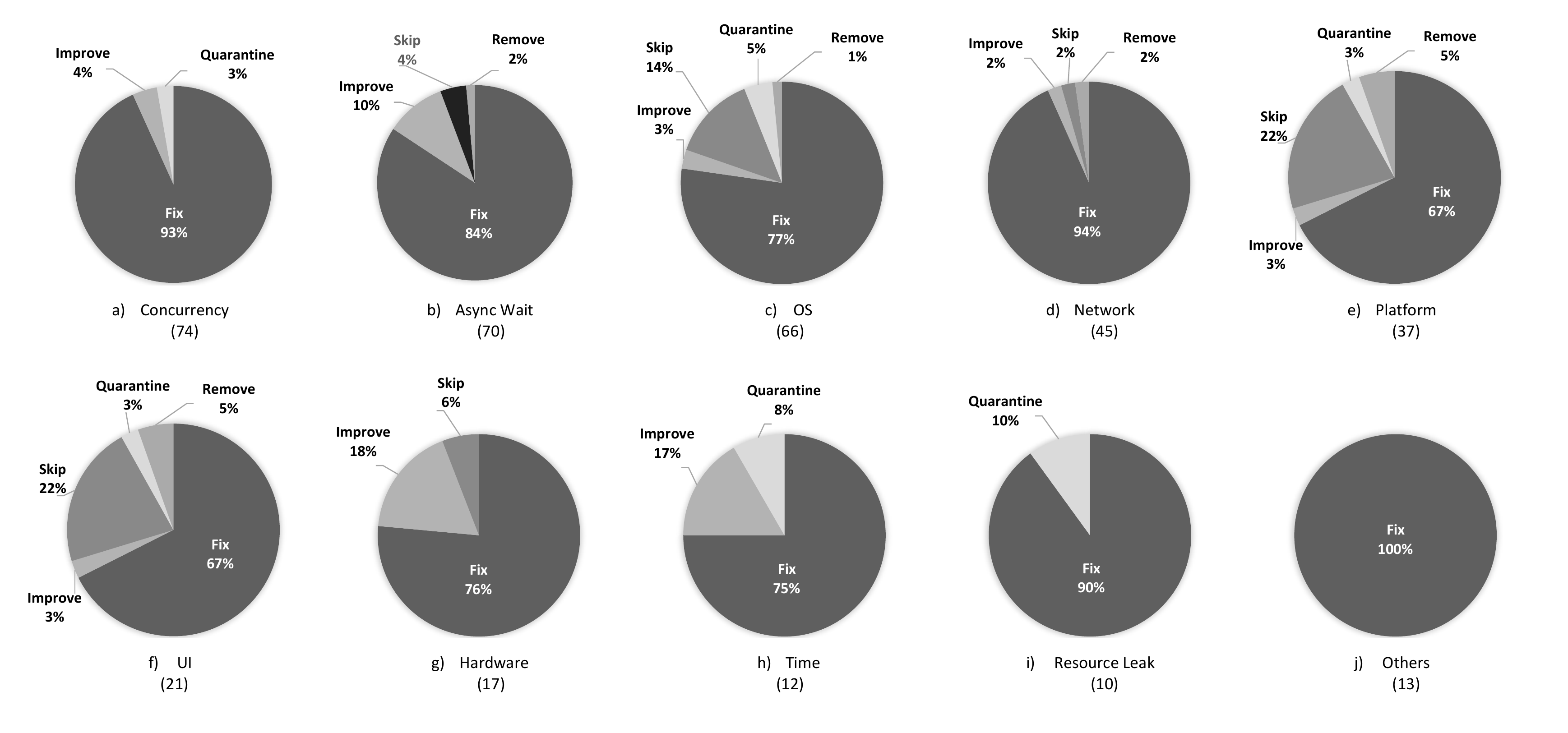}
\caption{Distribution of Flakiness Causes on Strategies (the number of commits are shown in parentheses)}
\label{fig:result}
\end{figure*}

Fig.~\ref{fig:result} demonstrates the percentages of strategies followed based on the cause of flakiness. The \emph{Fix} strategy represents the largest portion of each category. The most commits under \emph{Improve} category are related to \emph{UI}, \emph{hardware}, and \emph{time} causes. 
Similarly, 
as expected, most \emph{Skip} actions are taken to deal with \emph{OS} and \emph{Platform}-dependent  tests. The reason is that the tests are conditional in nature (will run in specific OSs or platforms), so that the test can run without any non-determinism. Also, the largest portion of \emph{Remove} is related to \emph{UI} and \emph{Platform} categories.
We list the most common change patterns used to fix the top four causes (concurrency, async wait, OS and network) in Table \ref{tab:change}.

\begin{table}[!htp]\centering
\caption{Common Changes for Four Top Categories}\label{tab:change}
\scriptsize
\begin{tabular}{lll}\toprule
Category &Common Change &Description \\\midrule
\multirow{6}{*}{Concurrency} &waitForEvent &\multirow{6}{*}{\parbox[t]{3cm}{In some cases, the test is flaky because of confliciting operations in different threads. By making the test wait and adding locks, they can prevent race conditions and synchronizing operations.}} \\
&increase timeout & \\
&setTimeout & \\
&setInterval & \\
&sleep & \\
&add lock & \\
&&\\
\midrule
\multirow{8}{*}{Async Wait} &waitForEvent &\multirow{8}{*}{\parbox[t]{3cm}{Most of the time, developers add time to the test to deal with async wait related test flakiness. For doing this, they usually increase time out or wait for special events.}} \\
&async & \\
&reordering code & \\
&Await & \\
&setTimeout & \\
&setInterval & \\
&increase timeout & \\
&timeoutPromise & \\
\midrule
\multirow{3}{*}{OS} &add condition &\parbox[t]{3cm}{In most of the cases, developers check the type and version and run/skip tests on specific ones.} \\
&async &\multirow{2}{*}{\parbox[t]{3cm}{Sometimes, the test is flaky because of the way the OS manages resources as it can take more time to run than expected.}} \\
&setTimeout & \\
&&\\
&&\\
&&\\
\midrule
\multirow{5}{*}{Network} &common.port &\multirow{5}{*}{\parbox{3cm}{In most cases, network related flakiness is caused by bad socket or port management, or lost internet connection.}}\\
&socket.setTimeout & \\
&socket.destory & \\
&socket.end() & \\
&client.shutdown() & \\

\bottomrule
\end{tabular}
\end{table}

\begin{qoutebox}{white}{}
\textbf{RQ2 findings:} We found that 82\% of flaky tests are fixed (to eliminate the flaky behaviour), 7\% improved (reduces flakiness), 7\% skipped or disabled, 2\% quarantined (for a later fix), and 2\% completely removed.
\end{qoutebox}

%% file: implications.tex
\subsection{Implications}

Our study has implications for both JavaScript testers (who write test code that might be flaky) and designers of flaky test detection and management tools. Our results show that JavaScript projects, as with projects in other languages studied to date, are prone to flaky tests, and that common causes of flaky tests in programs written using other languages (in particular Java and Python) are also prevalent in JavaScript.
Similarly to prior studies \cite{luo2014empirical,Thorve2018empirical}, we observed \emph{Concurrency} and \emph{Async Wait} as top causes for flakiness in JavaScript projects. 
Unlike in other empirical studies, \emph{Platform} is also a major cause of flakiness in JavaScript. This includes all scenarios where the test runs in a  different environment, browser, or OS. 
The implication from this observation is that implicit platform dependence (e.g. specific OS) must be made explicit in the test code.
Having automated ways of detecting platform related flaky tests would be of  great value to developers.
Any methods that can enhance developers’ understanding of the requirements and limitations of the different platforms used (on-the-fly) can potentially reduce the number of flaky tests caused by platform dependency.

The results of \textit{Concurrency} and \textit{Async Wait} being the most dominant cause of flakiness is expected given that JavaScript applications are highly asynchronous.
Unlike previous studies \cite{luo2014empirical,gruber2021empirical}, we observed a small number of flaky tests that were caused by \emph{Test Order Dependency}. The platform or testing frameworks are unlikely to be the cause for this low number, since testing frameworks for JavasScript (e.g. Jest\footnote{\url{https://jestjs.io/}}, Mocha\footnote{\url{https://mochajs.org/}}) have support for parallelising and ordering test execution and support for shared state across tests. It is possible that implicit shared state is more likely in object-oriented languages such as Java with static fields. Note that nearly 47\% of shared state in order-dependent tests \cite{parry2021survey} are external (e.g. files, database).

In addition, current detection tools focus mainly on languages like Java, Python and Ruby. Searching through the different resources and the recent literature surveys on flaky tests \cite{parry2021survey,zolfaghari2021root}, we found only one tool, NodeRacer \cite{endo2020noderacer}, that aims to manifest test flakiness causes by event races in JavaScript programs. There is related work on detecting other concurrency issues in JavaScript programs such as NRace \cite{chang2021race} or atomicity violations, NodeAV \cite{chang2019detecting}.
There is, generally, a lack of research on tools or techniques to deal with non-deterministic tests in JavaScript projects. More work is needed to provide detection tools that target other categories of causes for test flakiness in JavaScript, such as order dependent, OS and network flakiness.

%% file: threats.tex
\section{Threats to Validity}
\label{threatslab}


\noindent \textbf{Missing flaky tests related commits:}
We used a keyword-based approach to locate commits that are flakiness related. To provide an extended coverage, we used an extended list of 7 flaky test-related keywords in our search string (flaky, intermittent, fragile, brittle, flakey, non-deterministic test and intermit). It is still possible that we could miss either 1) commits that used keywords other than the one we identified in our research, or 2) flaky tests that have no associated commits (only reported in the issue tracker but have not been reviewed or actioned yet).

\noindent \textbf{Generalisation of the findings:}
The study considers commits from JavaScript projects that are publicly available on GitHub. The observations noted in this study, in terms of the categories of flaky tests, can be limited to the selected projects and may not be generalizable to other JavaScript programs. We mitigated this by selecting popular and highly starred projects that are managed by big communities and represent various types of applications (server, frontend/backend frameworks, web applications, graphics, etc.). Thus, our sample is somehow representative of real-world programs. Including additional projects will surely supplement our findings, and this may lead to more generalised conclusions.

\noindent \textbf{Manual classification of causes:}
A large part of the study depends on manual analysis of the data (commit messages and source code changes), which could affect the construct validity due to potential personal oversight and bias. In order to reduce potential false positives, all identified commits were independently classified by two authors (all commits were separately classified by the two of the authors), and when there were disagreements about a certain commit, we introduced a conflict resolution step where a third author was then involved to also independently classify the disagreed on commits. We followed that with a rigorous discussion of the causes between all authors until we reached an agreement (100\% agreement level). There were still issues with a number of commits that all authors agreed that the cause of flakiness is unknown or cannot be identified - those have been classified as ``hard to classify'' and then excluded from the analysis.

%% file: conclusion.tex
\section{Conclusion}
\label{conclusionlab}
In this paper, we investigate the presence of flaky tests in JavaScript projects. We first categorise flaky test-related commits based on the cause of flakiness, and then investigate the common strategies followed to deal with test flakiness.
Our study shows that the common causes of flaky tests in JavaScript are not different  from those noted in other languages - in particular, Java \cite{luo2014empirical} and Python \cite{gruber2021empirical}. We found that 70\% of flaky test commits in JavaScript are caused by one of the following: \textit{Concurrency}, \textit{Async Wait}, \textit{OS} or \textit{Network}. Unlike in previous studies \cite{luo2014empirical,gruber2021empirical},  which identify test-order dependency as one of the key causes of flakiness, we found only very few flaky test commits that are caused by test-order-dependency. In terms of fixing strategies, we note that the majority of flaky tests (82\%) are fixed to eliminate flaky behaviour completely. A smaller percentage of those flaky tests are either skipped (ignored), quarantined (to be fixed later, becoming a technical debt) or completely removed from the test suite.
These results can help future research on flaky tests, in particular  JavaScript tools designers in building tools that help to detect and remove flaky tests from test suites.